# New experimental findings of non-local transport in J-TEXT plasmas

Yuejiang Shi 石跃江[1*], Zhongyong Chen 陈忠勇[2], Zhoujun Yang 杨州军[2], Peng Shi 石鹏[2], Kaijun Zhao 赵开君[3], Partick H Diamond[3,4], JaeMin Kwon권재민[5], Wei Yan 严伟[2], Hao Zhou 周豪[2], Xiaoming Pan 潘晓明[2], Zhifeng Cheng 程芝峰[2], Zhiping Chen 陈志鹏[2], SeongMoo Yang 양성무[1], Chi Zhang 张弛[2], Da Li 李达[2], Yunbo Dong 董云波[3], Lu Wang 王璐[2], YongHua Ding 丁永华[2], Yunfeng Liang 梁云峰[2,6], SangHee Hahn한상희[5], HoGun Jhang 장호건[5], Yong-Su Na나용수[1]

[1]Department of Nuclear Engineering, Seoul National University, Seoul, Korea

[2]College of Electrical and Electronic Engineering, Huazhong University of Science and Technology, Wuhan, China

[3]Southwestern Institute of Physics, Chengdu, China

[4]CMTFO and CASS, University of California, San Diego, USA

[5]National Fusion Research Institute, Daejeon, Korea

[6]Forschungszentrum Jülich GmbH, Institut für Energie-und Klimaforschung-Plasmaphysik, Partner of the Trilateral Euregio Cluster (TEC), Jülich, Germany

[*]E-mail of corresponding author: yjshi@ipp.ac.cn   yjshi@snu.ac.kr

**Abstract**. In cold pulse experiments in J-TEXT, not only are rapid electron temperature increases in core observed, but also steep rises of inner density are found. Moreover, the core toroidal rotation also accelerates during the non-local transport process of electron temperature. These new findings of cold pulse experiments in J-TEXT suggest that turbulence spreading is possible mechanism for the non-local transport dynamics.

There are several outstanding experimental mysteries in magnetic fusion plasma research. A fast increase of the central electron temperature caused by the edge cooling in ohmic heating plasmas, the so-called non-local heat transport (NLT) [1] is one of these challenging issues. NLT was first observed in the TEXT tokamak 22 years ago, and in many fusion devices afterwards [2-12]. NLT phenomenon shows that the limitation of transport theory based on a local transport model. The underlying physics mechanism of NLT is still unclear.

One typical characteristic of NLT is this effect disappears with increasing electron density. The critical density where the NLT effect disappears is defined as cutoff density. The experimental results from C-mod [9, 10] and KSTAR [12] show the close correlation between the cutoff density of NLT and the plasma confinement status. In C-mod Ohmic heating plasmas [9,10], NLT can be achieved in linear Ohmic confinement (LOC) plasma with relative lower density and disappears in saturated Ohmic confinement (SOC) plasma with relative higher density. In KSTAR plasmas [12], the cutoff density of NLT can be can be significantly increased by electron cyclotron resonance heating because the transition density in ECH plasma from linear to saturated confinement is much higher than the density for LOC-SOC transition. The experimental behaviors of cutoff density of NLT looks consist with the some prominent theoretical hypotheses based on turbulence theory [13-17].

On the other hand, the latest experimental results in J-TEXT tokamak show a steep rises of core density and acceleration of core rotation, which are accompanied by rapid electron temperature increases. These simultaneous increments in three confinement channels are found for the first time in experimental fusion plasmas. The new experimental findings of non-local transport in J-TEXT in this paper provide a key clue to reveal more clear background physics mechanism for NLT. All the results presented in this paper are obtained from J-TEXT tokamak [18].

The major radius of J-TEXT is 1.05m. The minor radius of J-TEXT is 0.255m in this experiment. All the discharges are OH heating plasmas with circular limiter configuration. In J-TEXT, the electron temperature ($T_e$) is measured with the electron cyclotron emission radiometer (ECE) [19]. At the same time, the fluctuation of $T_e$ for core plasma is measured with correlation electron cyclotron emission radiometer (CECE) [20]. A 17-channel polarizer-interferometry system (POLARIS) [21] which covers the whole region of J-TEXT plasma from high field side to low field side, measures the electron density ($n_e$). The x-ray crystal spectrometer (XCS) [22] provides core toroidal rotation velocity. Multi-pulse supersonic molecular beam injection (SMBI) [23] is applied as cold pulse source to trigger NLT in J-TEXT [24]. The toroidal magnetic field ($B_T$) is fixed at 1.8T, which make the best spatial measured coverage of ECE.

Fig.1 shows the waveforms of a representative OH discharge with five SMBI cold pulse injection in J-TEXT. The pulse duration of SMBI is 0.3ms. The time interval between each SMBI pulse is 50ms i.e. 20Hz modulation frequency. The plasma current ($I_p$) and edge safety factor ($q_a$) are 150kA and about 3.7, respectively. The typical NLT effect in electron channel (core $T_e$ rise while edge $T_e$ drops) appears in every SMBI pulse, which is same as previous NLT experiments [1-12]. On the other hand, one notable point in fig.1 is the sawtooth period ($\tau_{sawtooth}$) also clearly increases for the all the 5 SMBI pulses. And the increasing pace of $\tau_{sawtooth}$ is coincident with the core $T_e$ rise.

There are still two plasma parameters can affect the $\tau_{sawtooth}$ of J-TEXT's plasmas with fixed $I_P$ and $B_T$. One is electron density, and the other is toroidal rotation velocity. Fig.2a shows the waveforms of density ramping-down discharge. $I_P$ and $B_T$ in fig.2 are same value as those in

fig.1. It can be seen in fig.2a that the $\tau_{sawtooth}$ decreases with decreased $\overline{n_e}$. On the other hand, the core $V_\phi$ also increases with density as shown in fig.2a. Here, we note that the core $V_\phi$ measured with XCS is always in counter-current direction in our experiments on J-TEXT. In this paper, the negative sign for $V_\phi$ or $\Delta V_\phi$ means that the rotation velocity or rotation acceleration is in counter-current direction. It is hard to say from fig.2a which one is dominant factor to affect $\tau_{sawtooth}$. Tangential neutral beam injection (NBI) is a very useful tool to actively control $V_\phi$. Although there is no NBI in J-TEXT at present, the resonance magnetic perturbation coils (RMP) in J-TEXT [25] is also a powerful actuator to change $V_\phi$. Fig2b shows the waveforms of one discharge with application of RMP at the same $B_T$ and $I_p$ as those in fig.2a and fig.1. It can be seen that the changing amplitude of $\overline{n_e}$ in fig.2b is much small than that in fig.2a. It is very clear in fig.3b that $\tau_{sawtooth}$ increase with $V_\phi$.

Although the effects of toroidal rotation on sawtooth period have also been reported both in experimental paper [26] and theoretical paper [27], we don't discuss which one ($n_e$ or $V_\phi$) is more dominant factor to affect $\tau_{sawtooth}$ in J-TEXT in this paper. We just simply assume both $\overline{n_e}$ and $V_\phi$ can affect $\tau_{sawtooth}$. Fig.3a and Fig.3b shows the scaling relation of $\overline{n_e}$ v.s. $\tau_{sawtooth}$ and $V_\phi$ v.s. $\tau_{sawtooth}$. It can be seen that $\tau_{sawtooth}$ has linear relation with $\overline{n_e}$ and 2nd order polynomial relation with $V_\phi$ within certain range. In fig.1, the increasing amplitude of $\tau_{sawtooth}$ ($\Delta\tau_{sawtooth}$) is about 0.25ms after SMBI injection. According to the scaling relation in Fig.2b, the increasing amplitude of ($\Delta\overline{n_e}$) should be about $0.22\times10^{19}m^{-3}$ to make corresponding $\Delta\tau_{sawtooth}$. The real experimental $\Delta\overline{n_e}$ after SMBI injection is only about $0.1\times10^{19}m^{-3}$, which is much lower than the estimated value from the scaling relation in Fig.3a. Moreover, $\tau_{sawtooth}$ already decreases while $\overline{n_e}$ still sustains high level for serval ten milliseconds. So the effect of $\overline{n_e}$ can be excluded for the prompt change of $\tau_{sawtooth}$ in fig.1. Now $V_\phi$ is the one possible parameter to make the rapid change of $\tau_{sawtooth}$ after SMBI injection in fig.1. According to the scaling relation in fig.3b, the increasing amplitude of $V_\phi$ should be around -1.4km/s to make corresponding $\Delta\tau_{sawtooth}$. Unfortunately, both XCS and visible active or passive spectrometer at present cannot catch the fast time scale of NLT and the small changing amplitude of $V_\phi$, simultaneously. More directly evidence of corresponding response of $V_\phi$ during NLT will be confirmed if the performance of related diagnostics can be greatly improved in the future.

The local density response during NLT is the other interesting point. Both time and spatial information of local electron density can be obtained from the 17-ch line integrated density of POLARIS with matrix inversion technique. The detail time evolution of the core density ($n_{e0}$) during the first SMBI pulse in fig.1 is shown in fig.4a. Fig.4b shows the evolution of profiles of $T_e$ and $n_e$ during the first SMBI pulse in fig.1. Obviously, there are two different time scale for the evolution of $n_{e0}$. One is the steep rise of $n_{e0}$ in serval milliseconds in the beginning. The other is the slow increase and sustainment in the following tens milliseconds. Moreover, the steep rise of $n_{e0}$ is synchronous with the fast increase of $T_{e0}$. For the J-TEXT SMBI experiments, the neutral penetration of SMBI mainly concentrated in the SOL and plasma edge region (r>23cm) [28]. If the steep rise of $n_{e0}$ is caused by local transport process, the averaged convection velocity from edge to core should be more than 40 m/s (~0.2m/5ms). On the other hand, the spatial particle transport coefficients based on perturbation principle also can be calculated for the 5-pulse SMBI plasma in fig.1 with modulation technology. The modulated amplitude and phase of local density are shown in fig.5a and fig.5b. The profiles of diffusion and convection velocity are shown in fig.5c and fig.5d. The convection velocity at r/a=0.75 reaches 90m/s. Normally, the particle convection velocity in other machines which are similar size as J-TEXT is around 1~10m/s , which is already much higher than the neo-classic prediction (abnormal transport). Compared to experiment results of other machines, both averaged 40m/s and localized 90m/s, which are estimated from local transport mode, are astounding high value. Similar to the fast rise of $T_{e0}$, the steep rise of $n_{e0}$ should be also caused by some non-local transport process.

Not only are the rapid electron temperature increases in core observed in J-TEXT cold pulse experiments, but also the steep rises of inner density and acceleration of core toroidal rotation are found. The J-TEXT results are the first experimental discovery of simultaneous fast NLT responses in multi-channels transport (electron temperature, particle, and momentum) in magnetic fusion plasma. Turbulence spreading [29-33] is the possible mechanism to explain the multi-channel NLT dynamics. The fast inward particle pinch induced by turbulence spreading [33] may explain the rapid increases of core density. The fast increases of electron temperature and acceleration of core rotation are also qualitatively predicted with the latest numerical simulation based on turbulence spreading theory [15].

The main idea to explain NLT dynamic in cold pulse experiment with turbulence spreading is as following. The turbulence is triggered by cold pulse in the outer unstable linear plasma and propagates to the core with very fast spreading process. The core part of LOC plasma is below marginal stability and provides room for further performance improvement until the SOC is reached. The fluctuation information of electron temperature of core region (r/a=0.18) from CECE for the shot in fig.1 is shown in fig.6. It can be seen in fig.6a that the low frequency fluctuation (f<10 kHz) of $T_e$ is dominated by some coherence mode. The relative high frequency fluctuation (20 kHz<f<100 kHz) with $f^{-1}$ dependence show the turbulence characters. The intensity of crosspower from 4 kHz to 10 kHz during NLT period clearly decreases, which can make improved confinement and increment of core $T_e$. On the other hand, the intensity of crosspower from 20 kHz to 60 kHz obviously increases during NLT period. The turbulence spreading maybe carried out with these relative high frequency fluctuations.

Although turbulence spreading can qualitatively explain the NLT dynamics, the quantitative simulation compared with real experimental data is necessary and important to enhance the credibility of related theoretical hypotheses.

**Acknowledgement**

This research is supported by the BK21 Plus project and R&D Programs through National Research Foundation of Korea (NRF) funded by the Ministry of Science, ICT and Future Planning of the Republic of Korea (No. 2014M1A7A1A03045368, and No. 21A20130012821). This research is also supported by National Magnetic Confinement Fusion Science Program funded by Ministry of Science and Technology of China (No. 2014GB108001, 2015GB111002, and 2015GB12003) and National Natural Science Foundation of China (No. 11775089).

**Figures**

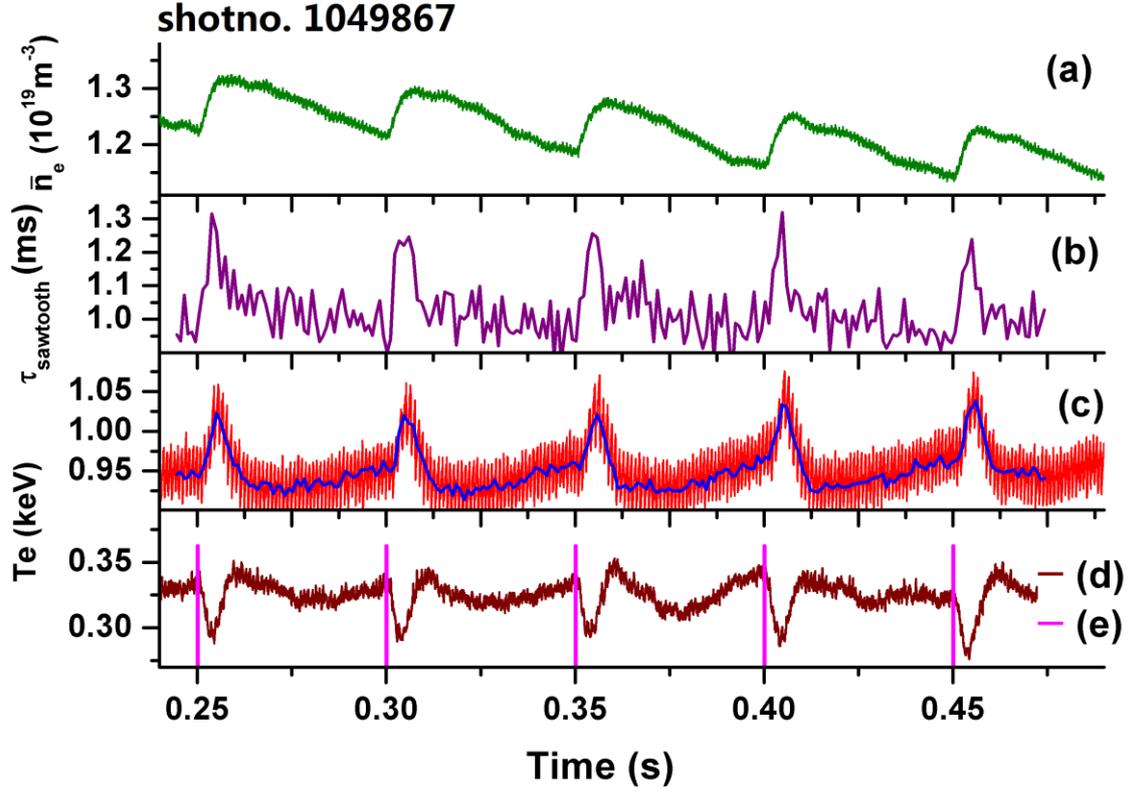

*Fig.1* The waveforms of cold pulse discharge with multi-pulse SMBI (shot no.1049867); (a) line-averaged electron density, (b) sawtooth period (c) core electron temperature, blue line is temperature at the middle time of each sawtooth cycle (d) edge electron density at r/a=0.79, and (e) SMBI pulse signal.

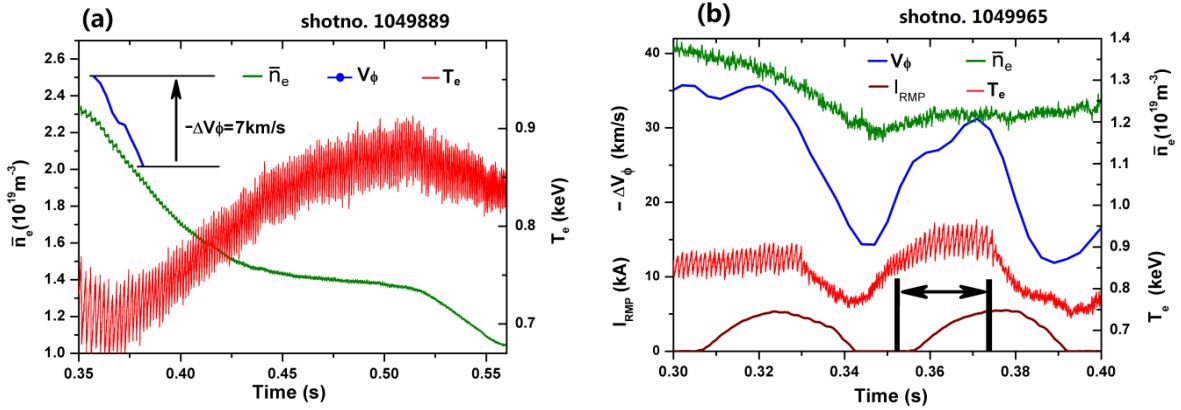

*Fig.2a* The waveforms of density ramping-down plasma (shot no.1049889);

*Fig.2b* The waveforms of RMP plasma (shot no.1049965). The error bar of $V_\phi$ is about 3km/s.

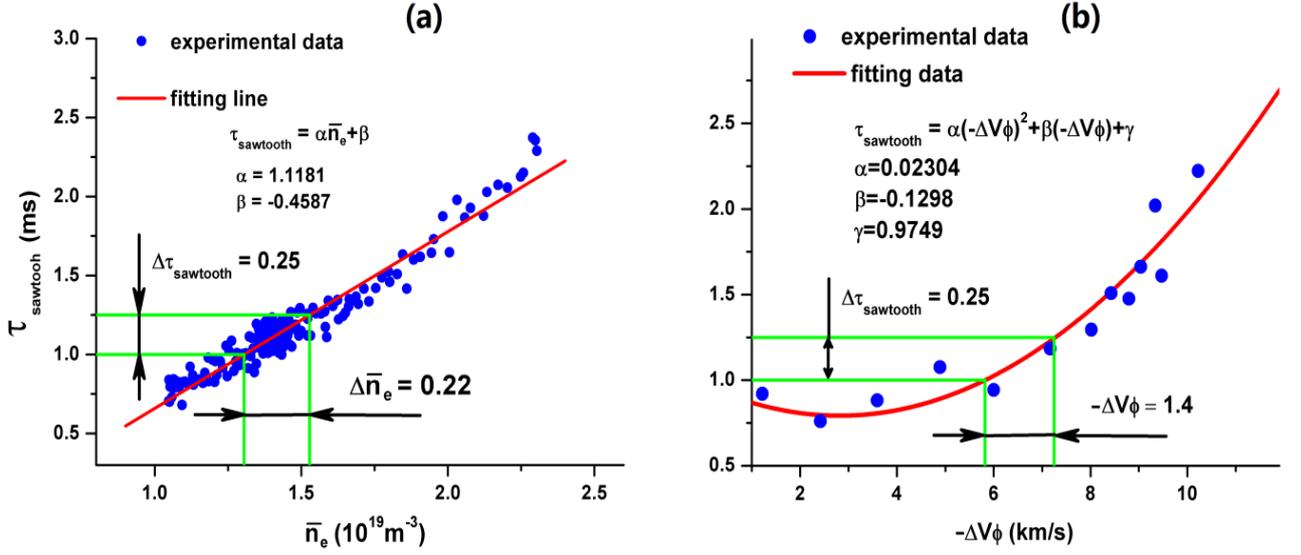

***Fig.3a*** *The scaling relation between $\overline{n_e}$ and $\tau_{sawtooth}$ from fig.2a;*

***Fig.3b*** *The scaling relation between $V_\phi$ v.s. $\tau_{sawtooth}$ from fig.2b.*

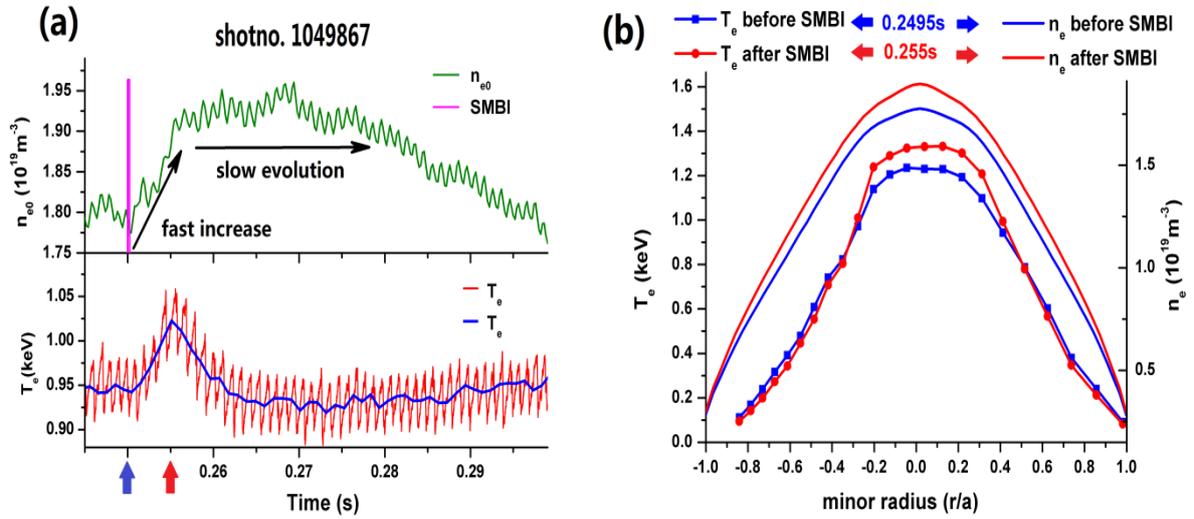

***Fig.4a*** *The detail time evolution of core $T_e$ and $n_e$ for the first SMBI pulse in fig.1. The blue line represents $T_e$ at the middle time of each sawtooth cycle*

***Fig.4b*** *The profiles of $T_e$ and $n_e$ for the first SMBI pulse in fig.1.*

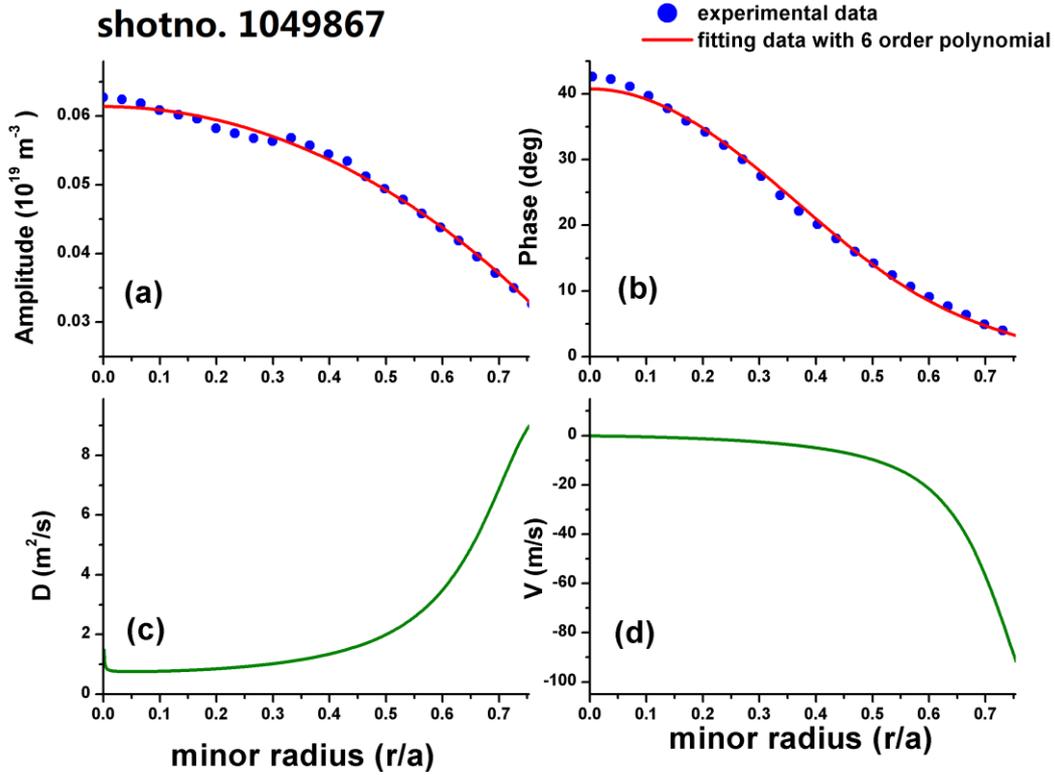

***Fig.5*** *Profiles of modulated amplitude (a) and phase (b) for the 5 SMBI pulses in fig.1*

*Profiles of diffusion coefficient (c) and convection velocity (d) calculated based on perturbation principle.*

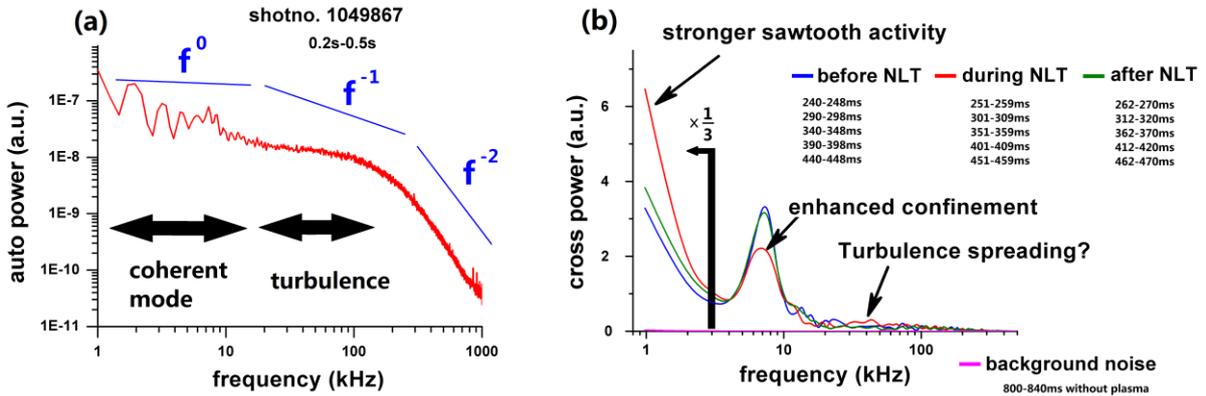

***Fig.6*** *(a) Autopower spectrum of one CECE channel located at r/a=0.18 for the shot in fig.1. The integrated time is 300ms. (b) Crosspower spectrum of two CECE channels at r/a =0.18. The integrated time for each crosspower spectrum is 40ms. The intensity of crosspower for f<3kHz is demagnified three times to make the behaviour of high frequency fluctuation more clear.*